\documentclass{aa}

\usepackage{graphicx}
\usepackage{txfonts}
\usepackage{natbib}
\usepackage{ulem}

\begin{document}

\title{Scales of the magnetic fields in the quiet Sun.}
\author{{\sc A.~L{\'o}pez Ariste}\inst{1} \and  A. Sainz Dalda \inst{2}}
\institute{THEMIS - CNRS UPS 853. C/ V\'{\i}a L\'actea s/n. 38200. La Laguna. Spain.  \and
Stanford-Lockheed Institute for Space Research, HEPL, Stanford University. Cypress Hall, 466 Via Ortega, Stanford, CA 94305. USA.
}
\offprints{Arturo.Lopez@themis.iac.es}
\date{Received ; accepted}
\begin{abstract} 
{The presence of a turbulent magnetic field in the quiet Sun has been unveiled observationally  using different techniques. The magnetic field is quasi-isotropic and
has field strengths weaker than 100G. It is pervasive and may host a local dynamo.}
{We aim to determine the length scale of the turbulent magnetic field in the quiet Sun.}
{ The Stokes V area asymmetry is sensitive to minute variations in the magnetic topology along the line of sight. Using data 
 provided by Hinode-SOT/SP instrument, we performed a statistical study of this quantity. We classified the different magnetic regimes and infer properties
of the turbulent magnetic regime. In particular we measured   the correlation length associated to these fields for the first time.}
{The histograms of Stokes V area asymmetries reveal three different regimes: one organized, quasi-vertical and strong field (flux tubes or other 
structures of the like); a strongly asymmetric group of profiles found around field concentrations; and a turbulent isotropic field. For the last,  we
confirm its isotropy and measure correlation lengths from hundreds of kilometers down to 10km, at which point we lost sensitivity. A crude attempt
to measure the power spectra of these turbulent fields is made.}
{In addition to confirming the existence of a turbulent field in the quiet Sun, we give further prove of its isotropy. We also measure correlation lengths down to
10km. The combined results show magnetic fields with a large 
span of length scales, as expected from  a turbulent cascade.}
\end{abstract}
\authorrunning{L\'opez Ariste and Sainz Dalda}
\titlerunning{Scales of the magnetic fields in the quiet Sun.}
\maketitle

\section{Introduction}
This work is dedicated to explore the properties of the turbulent magnetic field in the quiet Sun through the analysis of the asymmetries in the Stokes 
V profiles observed by Hinode-SOT/SP \citep{kosugi_hinode_2007,tsuneta_solar_2008} in a Zeeman-sensitive line. The existence of a magnetic field  turbulent in nature in those 
places with weak Zeeman signals and 
absence of temporal coherence in the plasma flows is taken for granted and, from Section 2 and thereafter, we shall not discuss whether these fields 
are turbulent or  not. Although our analysis provides further evidence of the existence of this turbulent field, we will assume that this existence is proven
and make our analysis in that framework, studying the coexistence of those turbulent fields with others more structured in nature, whose existence
is also beyond questioning. Despite that, but also because of that, we now dedicate a few lines to justify and put into 
context the turbulent nature of the magnetic field in most  of the quiet Sun.

The existence of a turbulent magnetic field accompanying the turbulent plasma in the quiet Sun is not a theoretical surprise, rather the opposite. Upon
the discovery of flux tubes in the photospheric network, \cite{parker_dynamics_1982} expressed surprise at the existence of these coherent magnetic structures
and referred to them as \textit{extraordinary state of the field}. He argued in that paper that only under conditions of temporal coherence of the plasma flows in 
the photosphere could these structures be stable. Consequently one could expect to find them in the photospheric network where advection flows concur
before dipping into the solar interior. Similar conditions can be found here and there in the internetwork in those intergranular lanes where plumes
have grown strong enough to survive granular lifetimes. Everywhere else the high Reynolds number of the photospheric plasma does not allow any
structure of any kind and a turbulent field, if anything,  is to be expected. Other theoretical analyses \citep{petrovay_turbulence_2001} confirm and 
insist on these turbulent fields. The first attempts of numerical simulations of magnetoconvection \citep{nordlund_dynamo_1992} revealed a magnetic
field whose field lines,  away from downflows, are twisted and folded even for the low Reynolds numbers (kinetic and magnetic) and for the
wrong ratio of both adimensional quantities. Independent of the existence of a local dynamo on which most of these simulations focus, the turbulent 
field is there.

Observationally, the picture has been different until recently. The discovery of flux tubes in the network \citep{stenflo_magnetic-field_1973} together
with the common observation of G-band bright points in high-resolution images of the photosphere has spread the idea that flux tubes are everywhere. Because
inversion
techniques for the measurement of the magnetic field mostly used Milne-Eddington atmospheric models that assume a single value of the magnetic field
per line of sight, whenever they have been used in the quiet Sun, a single field value was attached to a point in the photosphere, which additionally  spread the
impression that it was the field of the flux tube. Doubts were cast on this observational picture of the quiet Sun when magnetic measurements using
infrared lines produced  fields for the same points different than measurements with visible lines. The wavelength dependence of the Zeeman effect sufficed
to expose the fact that a single field could not be attached to a given point in the quiet Sun. Both measurements were right, in the sense at least
that they were measuring different aspects of the complex magnetic topology of the quiet Sun. Turbulent fields had in the meantime been the solution
offered by those observing the quiet Sun through the Hanle effect. The absence of Stokes U in those measurements and the high degree of depolarization
of the lines pointed to a turbulent field as the only scenario fitting their observations. Unfortunately,  the difficulties both in the observations 
 (very low spatial and temporal resolutions) and in the diagnostic (many subtle quantum effects involved) made  the comparison with the observations
using Zeeman effect difficult. 
The advent of the statistical analysis of Zeeman observations has solved the problem. First it was the observation that the quiet Sun, if one excludes
the network and strong magnetic patches from the data, looks suspiciously similar independent of the position on the solar disk that one is
observing \citep{martinez_gonzalez_near-ir_2008}. This independence of the measurements with the viewing angle pointed toward isotropy, a characteristic
of the quiet Sun fields later confirmed by \cite{asensio_ramos_evidence_2009}. Then came the realization that the average longitudinal flux density
measured in the quiet Sun at different spatial resolutions was roughly the same \citep{lites_characterization_2002,martinez_gonzalez_statistical_2010}.
This could only be interpreted that either the field was already resolved, but obviously it was not, or that the observed signals were just the result
of a random addition of many magnetic elements. Within the limit of large numbers the amplitude of this fluctuation only depends on the square root of 
the size and not linearly, as expected for a non-resolved flux tube. The turbulent field is in this way unveiled by the statistical analysis of 
Zeeman effect, and it was shown that Zeeman signatures in the quiet Sun were often merely statistical fluctuations of the turbulent field and not measurements
of the field itself \citep{lopez_ariste_turbulent_2007}.

The  solar magnetic turbulence was therefore explored in a statistical manner. Furthermore, it was explored assuming that different 
realizations of the magnetic probability distribution functions sit side by side. In this approximation one can compute the resulting polarization by just 
 adding up the individual contributions of each magnetic field. The Stokes V profile caused by the Zeeman effect of each individual magnetic field will
be  anti-symmetric with respect to the central wavelength, with one positive and one negative lobe. The areas of the  two lobes of every profile 
 are identical and their addition,
 the area asymmetry, will be zero. Adding many such polarization profiles will alter the resulting profile, but the area asymmetry of the final
profile will always be zero. A completely different result is obtained  if we consider the different realizations of the magnetic field probability distribution
function placed one after the other along the line of sight. Computing the resulting polarization profile now requires  integrating the radiative transfer 
equation for polarized light in a non-constant atmosphere. If the variations in the magnetic field along the line of sight are associated with velocities,
 the integration results in a profile that lacks any particular symmetry.

Therefore, measuring and analyzing the area asymmetry of the Stokes V profiles in the quiet Sun provides information on the properties of the turbulent 
magnetic field along the line of sight, in contrast with the previous studies, which only explored this turbulence in terms of accumulation of magnetic elements
in a plane perpendicular to the line of sight. At disk center, the line of sight means exploring those fields with depth, while near the limb it means 
exploring fields sitting side by side. Comparing asymmetries in statistical terms from quiet regions at different heliocentric angles provides us with
a probe on the angular dependence of the magnetic fields, and this is one of the purposes of this paper. In Section 2 we describe the asymmetries observed
by Hinode-SOT in these  terms and recover the three expected magnetic regimes: the structured and mostly vertical strong  fields (strong in terms of quiet 
Sun magnetism), the turbulent, ubiquitous, disorganized and weak fields and a class of profiles with strong asymmetries that can be observed at those places
where the line of sight crosses from one regime to the other, from turbulent to organized.

Focusing on these profiles assigned to turbulent magnetic fields, the value of the area asymmetry can be linked with the dominant scales of variation of the
magnetic field. The results on stochastic radiative transfer that allow us to make that link are recalled in section \ref{sec_scales}, in particular those of \cite{carroll_meso-structured_2007}. 
Thanks to those works we can quantitatively 
determine the correlation length of the magnetic field for every value of area asymmetry. From this determination we attempt to give an energy spectrum
for the magnetic turbulence at scales below the spatial resolution. For this attempt, we will use the longitudinal flux density as a lower boundary to the field strength,
and hence to the magnetic energy, and plot it versus the correlation length already determined. 
 The approximations and simplifications
made to reach this result may appear excessive to the reader. We argue that it is important not as an energy spectrum to be compared to numerical 
simulations
or to theoretical considerations, but as a first attempt that follows what we consider to be a promising tool and method for more elaborated and 
reliable determinations
of the magnetic energy spectrum. We also  stress  that through the proxy of the asymmetries of the profiles
 (seen through the models, tools, and results of stochastic radiative transfer), we can access 
  scales of variation of the magnetic field 
10 times smaller than the diffraction
limit of our best instruments and probably smaller  or comparable to the mean free path of the photons in the photosphere.

\section{Statistics of area asymmetries of the Stokes V profile.}

To collect data on asymmetries of the Stokes V profiles in the quiet Sun at different heliocentric angles,
we examined data from the SOT/SP instrument on board Hinode. 
Several large area scans of the quiet Sun at different positions on the solar disk were selected. Table \ref{latabla} summarizes the observational 
features of those data. The spatial sampling (0.15\arcsec $\times$ 0.16\arcsec) and spectral sampling (roughly 21 m\AA) were the same for all  
observing runs, but the exposure times changed from one to other. The data were calibrated using the standard procedure, which is available 
in {\it SolarSoft} and has been developed by B. Lites.
 The area asymmetry is measured as the integral of the profile over the Fe {\footnotesize I} 6302.5 \AA\ line, normalized to the total enclosed area, and with corrected 
polarity. 
Before this measurement a 
denoising  based on PCA (Principal Component Analysis) of the data has been performed \citep{martinez_gonzalez_pca_2008}. This allows one to establish true noise and 
signal levels for the signal.  In brief,
the eigenvectors of the correlation matrix of each data set  are computed at well-defined 
heliocentric angles. The data are reconstructed with only the ten first eigenprofiles. The rest are added to provide a measurement 
of noise. Histograms of this residual show the unmistakable Gaussian distribution shape of noise with typical values of $5\times10^{-4}$ the intensity of the 
continuum for the maximum of the distribution.

The first eigenprofile is, as usual in PCA techniques, the average eigenprofile and accordingly presents a nice antisymmetric shape with two well-defined lobes. 
This eigenprofile is used to automatically detect the polarity of every observed profile: a positive coefficient for this eigenprofile indicates 
a positive polarity profile.
 With this simple test we can automatically assign an unambiguous sign for the area asymmetry for all but a few anomalous cases, which can be easily 
disregarded because of their scarcity. The resulting sign of the area asymmetry is therefore related to the area of the blue lobe of Stokes V compared to that of the 
red lobe, and has no relationship with the polarity of the field. A positive area asymmetry results when the red lobe encloses a larger area than the blue lobe. 

Using a scalar parameter like the area asymmetry to describe the many different spectral features that a Stokes V profile can present can be perceived as simplistic
considering the magnetically and thermodynamically complex atmospheres in which it is formed. Studies and classifications of those shapes have been 
made and
interpreted in the past \citep{sanchez_almeida_observation_1992,viticchie_interpretation_2010} and we  refer to them for details on this aspect of
asymmetries. No doubt, a scalar like the
area asymmetry hides all that richness and we should worry about the possibility that our conclusions could be polluted or invalidated because of 
that reason. In face of that criticism we  claim  that the area asymmetry, contrary to the amplitude asymmetry, allows a relatively easy
analytical computation under unspecified variations of the atmospheric parameters   \citep{lopez_ariste_asymmetry_2002}. The shape of the profiles
is implicitly taken into account in  these computations and, however complex, it does not alter the known dependencies of
the area asymmetry on magnetic and velocity fields. Nevertheless,
the shape of the profile, or of any particular spectral feature, can be
attributed to particular variations in the atmosphere, something
that cannot be achieved with only a value for the area asymmetry.
As an example,  we  mention below the interpretation
of single-lobe V profiles as produced in atmospheres with
jumps along the line of sight. Because this work is limited to the
area asymmetry, we avoid any conclusion on particular conditions
along the line of sight. We will
therefore describe the atmospheres through a simple correlation length with no additional details on the geometry of the fields, details that the analysis of
profile shapes may eventually provide.                                                                                                  

\begin{center}
\begin{table*}[ht]
\caption[Table 1.] {Observational parameters of the data set. The spectral sampling was 21$m\AA$ for all observation runs.} 
\begin{center}
\begin{tabular}{ccccccc}
Date Time & X scale  & Y scale  & ( X, Y)  & $\mu$ & Exp. Time & S/N \\
   \multicolumn{1}{c}{\textbf{}} &
   \multicolumn{1}{c}{\textbf{(\arcsec)}} &
   \multicolumn{1}{c}{\textbf{(\arcsec)}} &
   \multicolumn{1}{c}{\textbf{(\arcsec, \arcsec) }} &
   \multicolumn{1}{c}{\textbf{}} &
   \multicolumn{1}{c}{\textbf{(s)}} &
   \multicolumn{1}{c}{\textbf{}} \\
\hline \\[-2ex]
2007-09-01 20:35 & 0.15 & 0.16 & (-153.1, 922.9) & 0.224 &  8.0 &   549. \\
2007-09-06 15:55 & 0.15 & 0.16 & ( -34.6,   7.0) & 0.999 &  8.0 &  1150. \\
2007-09-09 07:05 & 0.15 & 0.16 & ( 646.8,   7.2) & 0.739 &  9.6 &   957. \\
2007-09-27 01:01 & 0.15 & 0.16 &  (-1004.,   7.5) & 0.000 & 12.8 &   887. \\
\end{tabular}
\end{center}
\begin{center}
\label{latabla}
\end{center}
\end{table*}
\end{center}

\begin{figure*}[htbp]
\includegraphics[width=17cm]{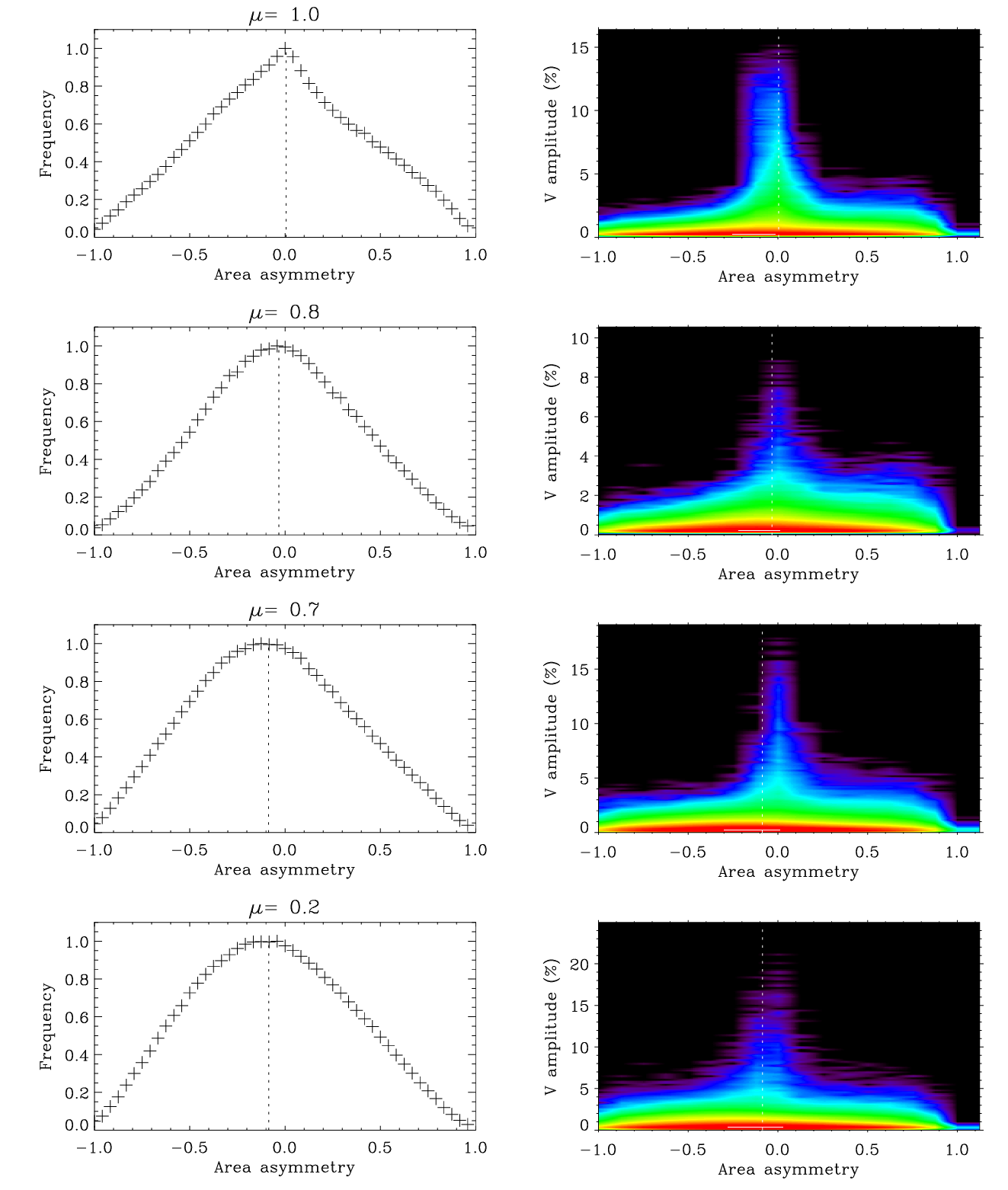} 
\caption{Histograms for four different heliocentric angles (from top to bottom) of the observed asymmetry in Stokes V profiles of the Fe \footnotesize{I}\ line at 6302.5 \AA. 
The left column presents the total histogram, while in the right column it is presented as a function of the amplitude of the profile in the ordinate
axis and with histogram values in color. For better visibility the logarithm of the actual value is shown. A vertical dashed line marks the position of the maximum of the
histogram in all plots.}
\label{view1}
\end{figure*}

Figure~\ref{view1} shows the data on area asymmetry for four different heliocentric angles ($\mu=1, 0.88, 0.7,$ and 0.2). The left column shows the
typical histogram of frequency as a function of signed area asymmetry. The position of the maxima of these histograms is given as a vertical dashed 
line. The histograms are quite common bell-shaped distributions with a slight bias toward negative area asymmetries, that is, toward profiles with
a blue lobe larger than the red one. That  bias  is stronger as we 
approach the solar limb \citep{martinez_pillet_active_1997}.  Larger blue lobes as observed have been observed in the past in active regions and network 
fields \citep{solanki_properties_1984}. They have been traditionally 
interpreted as being the result of the more usual gradients in the solar atmosphere \citep{solanki_can_1988}.
More precise information can be obtained if we make a 2D histogram as a function of the signal amplitude. This signal 
amplitude of the 
Stokes V profile in quiet Sun conditions can be safely interpreted as longitudinal flux density. Although given in terms of polarization levels, it
can as a rule of thumb be interpreted as $\times1000\ MW/cm^2$. These histograms are shown  in the right column of Fig. \ref{view1}.  In them
we plot in gray levels (color online) the histogram of area asymmetry of profiles with that signal amplitude. Indeed, to better display the 
strong, organized but relatively rare fields, the histograms are plotted not linearly but logarithmically. What is apparent in these plots is, first,
 a component of strong but weakly asymmetric fields with amplitudes higher than roughly 3\% (or $30\ MW/cm^2$). This class of fields is an important contribution to the 
histogram at
and near  disk center but is only marginally important near the limb. The obvious interpretation is that  these fields are the 
structured non-turbulent
and mostly vertical fields that can be found in the photospheric network or in particular intergranular lanes. This hypothesis can be confirmed 
by inspecting the actual magnetograms compared with the intensity maps (not shown here), rather than in the histograms as presented. Because they are mostly 
vertical, it is clear that their signature in Stokes
V will diminish as we approach the limb where these fields are seen transversally. Their asymmetries are small and centered around zero in all 
 data sets. This  indicates that these are very coherent structures with few variations (magnetic and velocity fields alike) along the 
line of sight.

\begin{figure*}[htbp]
\includegraphics[width=17cm]{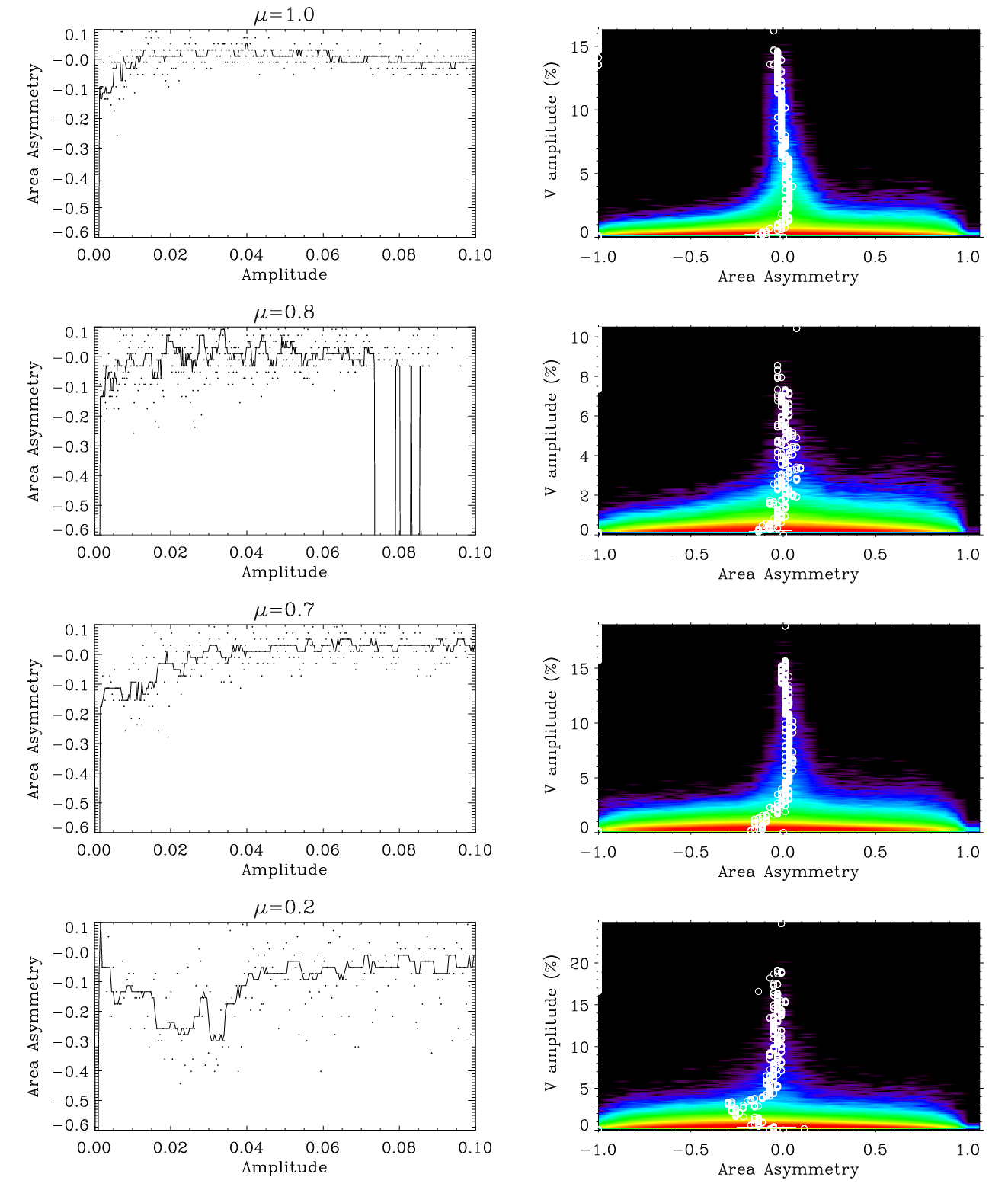} 
\caption{Average value of the area asymmetry of Stokes V for different heliocentric angles and as a function of the amplitude of the signal. In the
left column we present the actual measurements with a smoothed median line overplotted. In the right column this median line has been drawn over the 2D
histograms of Fig.1.}
\label{view3}
\end{figure*}

\begin{figure}[htbp]
\resizebox{9cm}{!}{\includegraphics{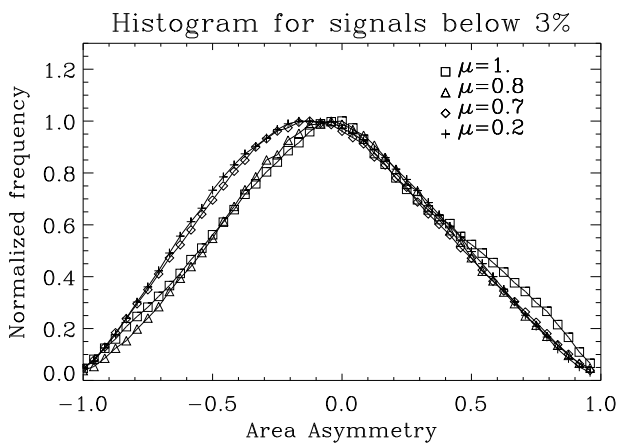}} 
\caption{Histograms of the asymmetries of the Stokes V profiles with amplitude below 3\% the intensity of the continuum. Four different heliocentric
angles are overplotted. The right slope of the histograms shows no dependence on the heliocentric angle, while in the left branch histograms from
profiles at less 
than 30 degrees ($\mu > 0.86$) show lower values. }
\label{view4}
\end{figure}

Except for this class of strong and weakly asymmetric profiles, the histograms are dominated by fields with low amplitude values 
(though well above the noise 
limit) that can produce almost any possible asymmetry from 0 to $\pm 1$. The most extreme asymmetries correspond to profiles with just one lobe, 
and we discuss them below. Now we concentrate on the distinction 
 between this  class of profiles with any asymmetry but weak amplitude  and the symmetric and strong-amplitude class described in the previous
paragraph. The separation between those two classes 
 can better be seen with the help of  Fig.\ref{view3}, particularly the right column of 2D histograms. We computed the position of the maximum of the
histogram for each value of the amplitude signal and plotted it versus amplitude in the left column of that figure. The strong fields have maxima at zero 
asymmetry. Maxima start drifting toward negative values for amplitudes below 5\% (or $50\ MW/cm^2$) and we can place a boundary between the two classes
at amplitudes of 3\% (or $30\ MW/cm^2$). 
 We observe that the fields responsible for this change in asymmetry
appear to be mostly insensitive to the heliocentric angle. This can be seen in Fig.\ref{view4} where the histograms were made exclusively with 
 profiles with amplitudes below the boundary of 3\% (or $30\ MW/cm^2$). We  overplotted in the same figure and at the same scale the histograms 
for all heliocentric angles to facilitate the comparison. The few differences  
with heliocentric angle that were seen in Fig.\ref{view1} have now almost
disappeared. Only a small variability in the slope toward negative area asymmetries is noticeable now. This variability is in contrast with the almost
perfect superposition of the histograms in their slope toward positive area asymmetries. We are unable to offer a definitive explanation for the 
variability in one of the slopes, but we nevertheless stress the weak dependence of those histograms on the  heliocentric angle. It is not proof, but 
yet suggestive evidence of the
general isotropy of those fields that their signature in terms of
area asymmetries is independent of the view angle under which
they are observed.

\begin{figure*}[htbp]
\includegraphics[width=17cm]{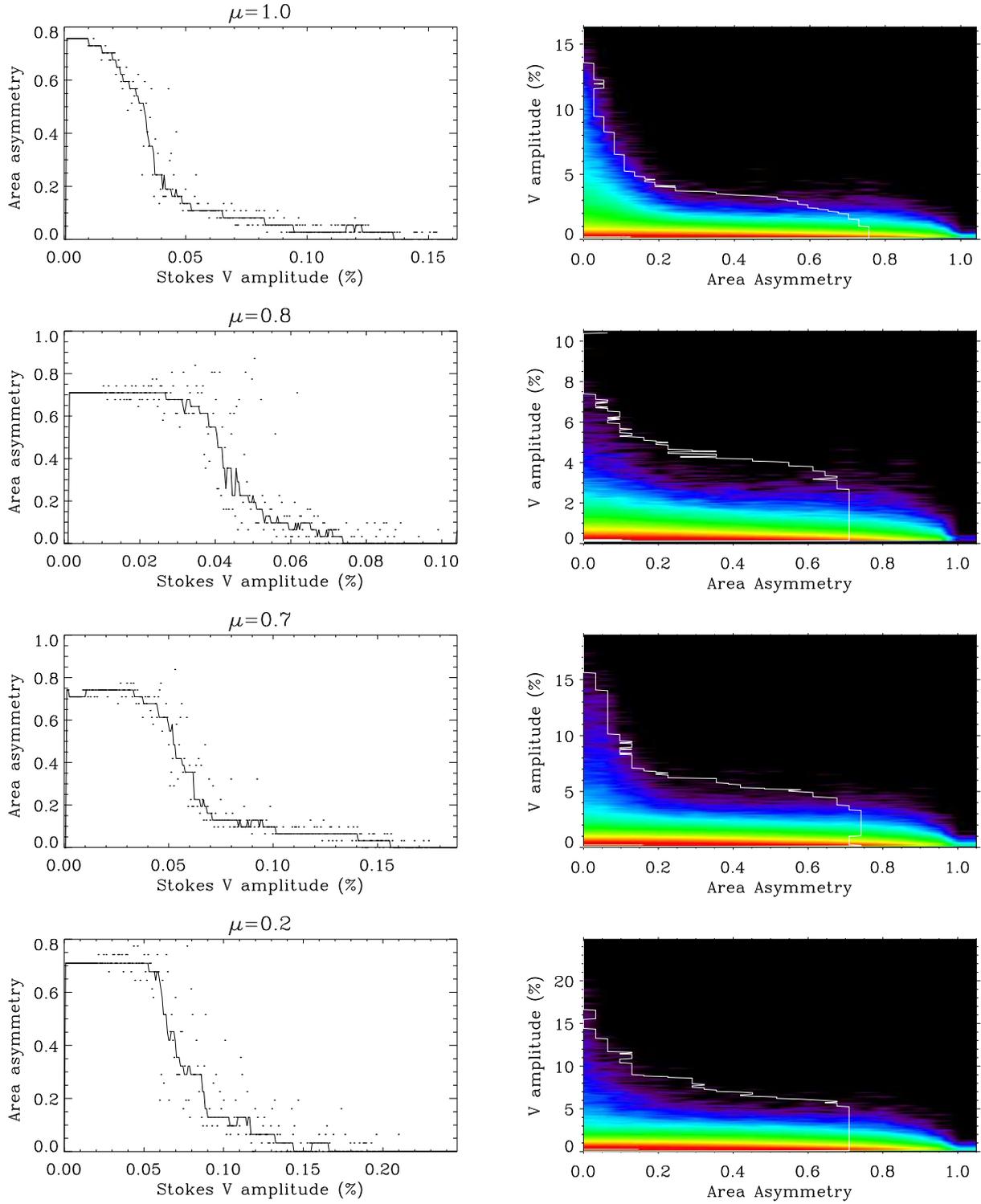}
\caption{95\% percentile value of the unsigned area asymmetry of Stokes V for different heliocentric angles and as a function of the amplitude of the signal. In the
left column we present the actual measurements with a smoothed median line overplotted. In the right column this median line has been drawn over the 2D
unsigned histograms of Fig.1.}
\label{view5}
\end{figure*}

It would be premature to identify any and all  fields with amplitudes below 3\% with the turbulent fields that pervade the quiet Sun. The histograms
of Fig.\ref{view4} suggest the conclusion that they appear to be 
isotropic,
which is an important characteristic of turbulent magnetic fields in the sense of fields frozen to a turbulent photospheric plasma at high values of the plasma $\beta$ 
parameter. The reason for this caution is that radiative transfer of polarized light through an atmosphere at the microturbulent limit concludes that
asymmetries larger on average than 0.53 are impossible \citep{carroll_meso-structured_2007}. 
Therefore, those observed profiles with amplitudes below 3\% but with extreme asymmetries cannot be considered as being formed
by radiative transfer through turbulent fields. At this point,
we recall  that those anomalous single-lobed profiles, with extreme asymmetries, are not observed at random over the solar surface but are instead  
systematically observed
over the boundaries of regions with concentrations of strong magnetic fields \citep{SainzDalda12}. This realization suggests that the origin of those 
profiles is a line of sight that
crosses from a weak disordered magnetic region to a strong and organized magnetic structure, with magnetic and velocity fields completely uncorrelated 
in one and the other regions. Simulation of these scenarios has successfully reproduced these single-lobed profiles with strong area asymmetries.
We should therefore distinguish among two different magnetic scenarios: the turbulent field  we are interested in studying and the single-lobed Stokes V profiles
appearing at the boundaries of concentrations with strong magnetic field. It is essential to be able to distinguish between these two classes of profiles by looking at the
statistics of the asymmetries alone. 
Consequently, in Fig.\ref{view5}, we try to ascertain the contribution of single lobe profiles to our histograms.
Since it is difficult to see the true weight of profiles with a given asymmetry in a histogram with a logarithmic scaling, we  computed the 
95 percentile of the area 
asymmetry, that is, the value of the asymmetry such that 95\% of the profiles with the same amplitude have an area asymmetry smaller than, or equal to, that 
value. We see in the left column of Fig.\ref{view5} that the 95 percentile area asymmetry is almost constant for all these fields with amplitudes
below 3\%. That is, once we enter into the class of weak fields, profiles with extreme asymmetries beyond 0.7 contribute to the histograms with less 
than 5\% of the cases, and that is independent of amplitude or heliocentric angle. We consider that given the measurement conditions and the rough
proxy that the 95 percentile is, the constant 
value of 0.7 is to be identified as the microturbulent
asymmetry limit.  We conclude therefore that 95\% of the profiles with weak amplitudes have asymmetries that can be explained as radiative transfer of 
polarized light through a turbulent magnetic field.  The constancy of this value for any amplitude and any heliocentric angle 
is in our view a strong statement in favor of this identification. What
other explanation as simple as the one we put forward can be offered to explain that asymmetry values fall predominantly in the range (0-0.7) 
independent of both signal amplitude (i.e. longitudinal flux density) and heliocentric angle?

Summarizing, in this section we have successfully identified  three different magnetic regimes in the data of area asymmetries: first, a strong, coherent, 
organized, mostly vertical and weakly asymmetric field, second, the anomalous single-lobed Stokes V profiles with strong asymmetries contributing to 
less than 5\% of the cases, which are 
located in the boundaries of strong-magnetic field concentrations, and finally a turbulent field that is quasi-isotropic and has asymmetries in 
the range expected from stochastic radiative transfer calculations. With this classification we reach the first and primary result of the present work,
 which is that also through area asymmetries we identify the presence of a turbulent, quasi-isotropic magnetic field component in the quiet Sun.
Asymmetries are sensitive to the
variations of the magnetic and velocity field along the line of
sight. This is complementary to previous studies that concentrated
on statistics of amplitudes and amplitude ratios that were
sensitive to magnetic fields placed side by side over the photosphere.
Despite this complementarity, we arrive at the same
conclusion of the presence of a turbulent and isotropic field in
the quiet Sun.

\section{Scales of the turbulent magnetic field in the quiet Sun.}\label{sec_scales}

We now focus on the regime of turbulent fields identified in the statistical data of area asymmetries. These  fields are mostly isotropic, as  is
to be expected from turbulence associated to the photospheric plasma, and they have low amplitudes. They should not be interpreted as
the direct, which is best described by a probability distribution function \citep{trujillo_bueno_substantial_2004,dominguez_cerdena_distribution_2006,
lopez_ariste_turbulent_2007,sanchez_almeida_simple_2007,sampoorna_zeeman_2008,stenflo_distribution_2010}. A good choice for that distribution function is 
a Maxwellian
for the modulus of the vector (the field strength) \citep{dominguez_cerdena_distribution_2006,lopez_ariste_turbulent_2007,sanchez_almeida_simple_2007}, which is 
 fully isotropic for its inclination and azimuth in whatever reference system of choice. The
average value of this probability distribution function for a vector field is the null vector. But it would be a mistake to interpret from that zero average that 
no polarization  signals would be observed
if this were the topology of the field in the quiet Sun. The average of the distribution is only attained at the limit of infinite realizations. Only
 if the scale of variation of the magnetic field, along the line of sight or across our resolution element, were zero  the average would be 
realized and observed. 
But in reality that scale of variation is nonzero, and both along the formation region of the observed spectral line and across the spatial resolution 
element of our 
observation, a finite number of magnetic field realizations is  found. Their integration will fluctuate around the average zero value, but will not
be zero. Those fluctuations are what is to be expected to become the observed of longitudinal flux density in a turbulent magnetic field scenario. The variance
of those fluctuations will diminish with the square root of the number of realizations. Therefore one should expect in the approximation of magnetic realizations sitting side by side 
across the spatial resolution element  that as the spatial resolution increases, the average longitudinal flux increases not with
the square of the size of the resolution element (as could be expected if unresolved magnetic structures were present), but linearly with that size.
This is what is indeed observed in the quiet Sun, where observations of the average flux density in the quiet Sun have shown no particular difference
between instruments with 1\arcsec, 0.6\arcsec\, or 0.3\arcsec\ resolutions \citep{martinez_gonzalez_statistical_2010}. This is one of the strongest arguments in favor of 
a turbulent magnetic field in
the quiet Sun. 
\begin{figure}[ht]
\resizebox{9cm}{!}{\includegraphics{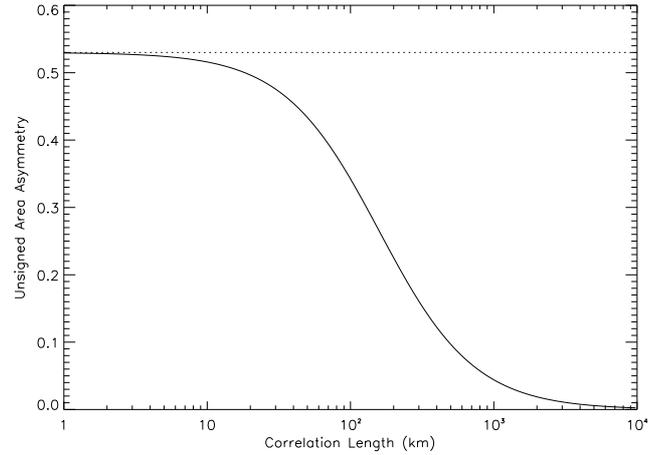}} 
\caption{Absolute value area asymmetry as a function of the correlation length obtained after resolving the radiative transfer equation of polarized light through stochastic atmospheres. The
dashed line at $\sim 53\%$ (or 0.53)   marks the microturbulent limit for asymmetry parameters caculated under the  MISMA approximation (data from \cite{carroll_meso-structured_2007}, right panel 
of Fig. 1).}
\label{curva}
\end{figure}

In terms of area asymmetries we are concerned about the number of realizations of the magnetic probability distribution function along the formation
region of the observed line. Contrary to the case of the spatial resolution, it is not evident to change the size of the sampling region. The formation region
of a spectral line is what it is and cannot be changed  
and, in the photosphere, most spectral lines of interest have formation regions that are too 
similar in size and position to be of any interest in the present terms. However, it is of great advantage that the size of those formation regions
is quite small, a few hundred kilometers at most, and the response functions over those regions are not flat but present a shape that has led many authors
to speak of height of formation rather than region of formation despite warnings about this oversimplification \citep{sanchez_almeida_heights_1996}.
 Therefore, when observing area asymmetries we are sampling those small regions of formation. We are directly sampling scales of less than 100 $km$.
These scales  are difficult to access for any diffraction-limited imaging instrument. Therefore area asymmetries present a clear advantage in terms of sensing the
scales of variation of turbulent magnetic fields.

To correctly interpret the actual values of area asymmetries in terms of scales of variation of the magnetic field, we require  a theory for 
radiative transfer of polarized light through stochastic atmospheres. Such models can be found in the works of \cite{auvergne_spectral_1973,frisch_non-lte_1976,frisch_stochastic_2005,frisch_stochastic_2006}
and \cite{carroll_line_2005}, to give a few solar-related examples. Here we 
used the result of \cite{carroll_meso-structured_2007} since they explicitly computed area asymmetries as a function of the correlation length of the 
parameters in their stochastic
atmospheres. 
The correlation length, the key concept for the goal of this paper, describes the distance between two points along the light path for which the probability of the magnetic
field being the same is small \footnote{In the particular model adopted by \cite{carroll_meso-structured_2007}, this probability is  $1/e$.}, assuming a Markovian model
in which this probability falls with increasing distance. This correlation length describes \textit{``the mean length scale of the structures''}, to quote 
 \cite{carroll_meso-structured_2007}. Roughly in every correlation length along the light path there is a change in the value of the atmospheric parameters, and gradients 
consequently appear producing asymmetries in the profiles. If the correlation length is longer than the formation region of the line, there will be no change in parameters, the magnetic and velocity fields
will be constant and no asymmetries must be expected. With a shorter correlation length than the formation region, more gradients appear and accordingly asymmetry grows.
 However, if the correlation length shrinks beyond the mean free path of the photon in the atmosphere, there are no atom-photon interactions to take 
those small-scale gradients into account  and the asymmetries saturate to a \textit{microturbulent} limit. This is the expected dependence of asymmetries with correlation length; and this
is what \cite{carroll_meso-structured_2007} found in their calculations, shown in their Fig. 1 (right side),
 which we reproduce here in Fig. \ref{curva} for the sake of completeness. 
The details on the computation of the figure are given by the authors of the referred work. The figure refers to the same \ion{Fe}{I} line as used in the observations
of the present work. 
Using the computed relation
shown in that figure but reversing the argument, the observed asymmetries of those profiles (identified as belonging to the turbulent regime in the previous section)
 can be translated
into correlation lengths. Since all asymmetries from 0 to the micro turbulent limit of 0.53 are observed, we can already conclude that in the data
considered for this work there are profiles formed in regions with variations
of the magnetic field at scales of 10km. Below that length the asymmetries quickly and asymptotically approach the microturbulent limit and we loose
our sensitivity. It is important, before proceeding any further, to stress that point: the observation of asymmetries in the Stokes V profiles allows
us to identify tiny scales of variation for the magnetic fields that produce those signals. Many of the strong signals identified in the previous section 
with the structured and mostly vertical fields presented asymmetries around zero. That is, those structures had correlation lengths or were coherent
over scales of hundreds of km up to 1000km. This is what we should expect from structured fields and it is justified that the best instruments 
in terms of spatial resolution start resolving them more and more frequently. 

On the other hand, for the fields that we assigned to the turbulent regime,  the range of variation of observed asymmetries fills the allowed range
of variation all the way up to the microturbulent limit. Thus we have identified profiles that arise from regions with magnetic fields varying at scales
of less than 100km and eventually down to 10km. These are actual observations, not simulations or estimates. To put those scales 
in another context, they correspond to the 
diffraction limit of an instrument with 7 meters of diameter of entrance pupil. More interesting, it is comparable to, or smaller than, the mean free 
path of the photon \citep{mihalas_stellar_1978,sanchez_almeida_heights_1996}.
We emphasize these comparisons because they show the finesse of the diagnostic that is accessible through the asymmetries of the Stokes V
profile.

After translating the  asymmetry of each profile belonging to the turbulent regime to correlation lengths, we noticed that we also have  a measurement
of the longitudinal flux density. It is tempting to try and  convert it into magnetic energy which would allow us to produce the all-important magnetic energy
spectrum of the turbulent magnetic field \citep{nakagawa_energy_1973,knobloch_spectrum_1981}. 
\begin{figure}[ht]
\resizebox{9cm}{!}{\includegraphics{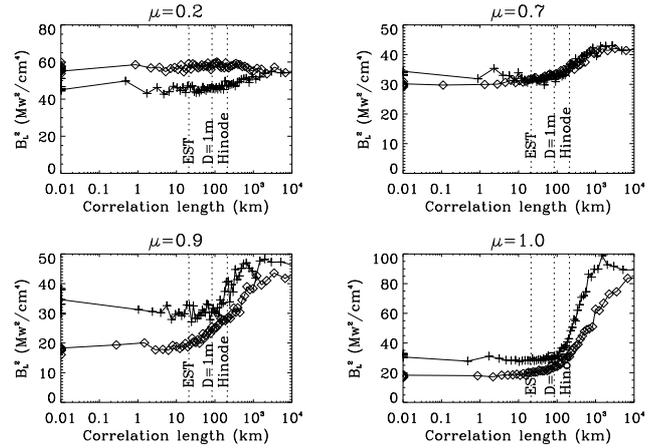}} 
\caption{Estimates of the magnetic energy spectrum of the turbulent magnetic field of the quiet Sun observed through the area asymmetries of Stokes V. For
each heliocentric angle, the two curves correspond to positive (crosses) and negative (diamonds) asymmetries. The diffraction limits of three characteristic
solar instruments are given for reference.}
\label{view6}
\end{figure}

There is a caveat attached to this, however.  It is a main result of this work to have identified the three different regimes as seen from the point of view of area 
asymmetries in the Stokes V profiles. For the turbulent regime, the actual asymmetry obviously depends on the correlation length of the parameters, including
the magnetic and velocity fields, of the stochastic atmosphere used in the model. The details of the stochastic radiative transfer may be a matter of 
discussion and study, but the general relation of asymmetries with correlation length is not in doubt. Our claim to have measured magnetic fields
varying at scales below 100 $km$ is also a robust result of this work. To proceed to a magnetic energy spectrum requires a series of assumptions, 
approximations and simplifications, which may appear as too strong. We nevertheless proceed because 1) those simplifications and approximations
are, in our opinion, still justified as fairly educated guesses, and 2) because the final result, even if crude, still illustrates how powerful a tool
the analysis and study  of area asymmetries can be.

With those cautions, we assume that the measured longitudinal flux density is a fair proxy of the average field strength of the distribution of
fields  present along the line of sight. Because of the projection along the line of sight we can assume that it is indeed a
lower bound to that field strength. We furthermore assume that the probability distribution function underlying the observed distribution is Maxwellian
for the field strength. This distribution is fully determined by the position of its unique maximum. Our measurement of the longitudinal flux density 
is seen as a lower boundary proxy to that maximum. The magnetic energy of the Maxwellian is computed as
$$
E=\frac{\int B^2 p(B)dB}{\int p(B)dB},
$$
where $p(B)$ is the Maxwellian probability distribution function.
The integrand looks in shape very much like the original Maxwellian, but its maximum is shifted toward higher values and the high-field wing is emphasized. Because
of these similarities we can claim that the integrand is fully determined by the new maximum of which the maximum of the original Maxwellian is a lower 
boundary. Our measurement of the longitudinal flux density therefore provides a lower limit to the magnetic energy in the distribution
of fields along the line of sight in our turbulent atmosphere.
In Fig. \ref{view6} we plot at four different heliocentric angles this magnetic energy  as a function of the correlation length. They are in
a sense crude  approximations to the magnetic energy spectrum with sensitivities down to scales of 10 $km$.  

Owing to the approximations and simplifications made it is difficult to extract much information from those energy spectra. 
Clearly, where the strong and structured fields are seen, near disk center, they dominate the energy spectrum and create a bump at the larger scales and a slope
toward smaller scales that can be compared to previous results \citep{nakagawa_energy_1973}. But that appears to be a valid description only for 
the more organized and mostly vertical fields. For the fields that we included in the turbulent regime,  the
spectrum appears to be decreasing over all the scales at which we are sensitive with a tendency to flatten out toward the 10km limit and certainly below 
that limit, where we are not sensitive. 

\section{Conclusion}
We studied the properties of the turbulent magnetic field that we assumed to pervade most of the quiet Sun. It is not the purpose 
of this work to demonstrate the existence of this turbulent component. Many observational and theoretical arguments are accumulating to prove its existence,
 one of which is the way in which we are detecting it through Zeeman polarimetry. We added new observational arguments confirming its
existence but, beyond that, we  studied it through
the asymmetries in the Stokes V. We have collected and analyzed the asymmetries of the Stokes V profile of the Fe \textbf{\footnotesize{I}} line at
 6302.5 \AA\  observed with Hinode-SOT/SP 
at several heliocentric angles. The histograms of those asymmetries for different signal amplitudes (or longitudinal flux densities) reveal three different
magnetic regimes. We notice that although in our results we talk about three regimes as separated, this is only a classification to facilitate the
 understanding of their existence.
 Needless to say, these regimes are simultaneously taking place in the quiet Sun and they are not uniquely separated by the area asymmetry 
value but also by their flux density or magnetic energy spectrum.
The first one is made of strong signals with weak area asymmetries that are mostly distributed around zero. They are mostly vertical, as can
be concluded from the field's importance at disk center and its waning as we approach the limb. The weak asymmetries, when interpreted in terms of magnetic
and velocity field variations along the line of sight, are a signature of big correlation lengths, that is, these structures present coherent fields
all throughout them, which justifies referring to them as magnetic structures. 

The second magnetic regime detected is characterized by its weak longitudinal flux density (below $30 MW/cm^2$ at Hinode-SOT/SP spatial resolutions)
and  unsigned area asymmetries that span the full range from 0 through the observed microturbulent limit of 0.7. This is slightly higher than the 
theoretically predicted value of 0.53. We identified them as the turbulent field that pervades
the quiet Sun. But before focusing on this, we first turn to the third magnetic regime detected, also made of weak signals, but with strong 
asymmetries beyond the microturbulent limit. Distinguishing between these two regimes is a delicate matter. It is done through two independent 
observations.
The first is that these single-lobed Stokes V profiles with area asymmetries beyond 0.6 or 0.7 are to be found mainly in the boundaries of 
strong magnetic field
concentrations. This observation has prompted us to interpret  them as profiles arising from lines of sight that cross from a strong magnetic structure
to the turbulent quiet Sun. These jumps in the magnetic and thermodynamic parameters of the atmosphere along the line of sight can produce  anomalous profiles like this.
The second observation to tell them apart from the turbulent regime is the 95 percentile of the asymmetry histogram. This statistical measure is in the data independent
of the amplitude signal or heliocentric angle and roughly equal to the microturbulent limit. The independence of the amplitude signal is the hardest test that makes us confident
that the two regimes can be  also directly separated in the histograms.

As a first result of this work we identified the magnetic turbulent fields in the data of area asymmetries in the Stokes V profiles. 
We furthermore realized that the histograms
of the asymmetries of these profiles are almost independent of the heliocentric angle, which is another observational result supporting the
isotropy of these fields.
 Next we used the results of radiative transfer of polarized light through stochastic atmospheres, particularly the work of 
 \cite{carroll_meso-structured_2007} to translate  those asymmetries into correlation lengths for the atmospheric parameters, specifically the 
magnetic and velocity fields.
We stressed the sensitivity to small scales that area asymmetries provide. The formation region of the observed spectral lines spans a few hundred kilometers
and  is mostly concentrated in a few tens of kilometers. This is much better than the spatial resolutions achievable through direct imaging. 
We emphasized that we measured scales of variation and not physical structures. Those scales of variation are the best description of a turbulent
field, without identifiable structures over the range of scales corresponding to the inertial regime of the turbulence. The smallest of these scales 
is the  scale of dissipation, which has been estimated to be as small as 100m  \citep{pietarila_graham_turbulent_2009}. Our measurements appear to identify 
magnetic scales that are still short of this dissipation scale by a factor 100.

With a powerful tool like this we determined the range of scales of variation of the turbulent magnetic field and saw that examples of all scales are found:
from the hundreds of kilometers of magnetic concentrations that other instruments start to resolve through direct imaging through the tiniest scales
of 10km at which area asymmetries loose their sensitivity.  In an effort to exploit this information on scales, we attempted to measure the magnetic energy
through the proxy of the longitudinal flux densities. There are many approximations and simplifications, but beyond the confidence on the final result
we  stress again the potential of the tool if only better determinations of the magnetic energy can be used. 

\begin{acknowledgements}
Hinode is a Japanese mission developed by
ISAS/JAXA, with NAOJ as domestic partner and NASA and STFC(UK) as international partners. It is operated in cooperation with ESA
and NSC (Norway). The Hinode project at Stanford and Lockheed is
supported by NASA contract NNM07AA01C (MSFC).
\end{acknowledgements}

\bibliographystyle{aa}

\begin{thebibliography}{35}
\expandafter\ifx\csname natexlab\endcsname\relax\def\natexlab#1{#1}\fi

\bibitem[{{Asensio Ramos}(2009)}]{asensio_ramos_evidence_2009}
{Asensio Ramos}, A. 2009, The Astrophysical Journal, 701, 1032

\bibitem[{Auvergne {et~al.}(1973)Auvergne, Frisch, Frisch, Froeschle, \&
  Pouquet}]{auvergne_spectral_1973}
Auvergne, M., Frisch, H., Frisch, U., Froeschle, C., \& Pouquet, A. 1973,
  Astronomy and Astrophysics, 29, 93

\bibitem[{Carroll \& Kopf(2007)}]{carroll_meso-structured_2007}
Carroll, T.~A. \& Kopf, M. 2007, Astronomy and Astrophysics, 468, 323

\bibitem[{Carroll \& Staude(2005)}]{carroll_line_2005}
Carroll, T.~A. \& Staude, J. 2005, Astronomische Nachrichten, 326, 296

\bibitem[{{Dom{\'i}nguez Cerde{\~n}a} {et~al.}(2006){Dom{\'i}nguez
  Cerde{\~n}a}, {S{\'a}nchez Almeida}, \&
  Kneer}]{dominguez_cerdena_distribution_2006}
{Dom{\'i}nguez Cerde{\~n}a}, I., {S{\'a}nchez Almeida}, J., \& Kneer, F. 2006,
  The Astrophysical Journal, 636, 496

\bibitem[{Frisch \& Frisch(1976)}]{frisch_non-lte_1976}
Frisch, H. \& Frisch, U. 1976, Monthly Notices of the Royal Astronomical
  Society, 175, 157

\bibitem[{Frisch {et~al.}(2005)Frisch, Sampoorna, \&
  Nagendra}]{frisch_stochastic_2005}
Frisch, H., Sampoorna, M., \& Nagendra, K.~N. 2005, Astronomy and Astrophysics,
  442, 11

\bibitem[{Frisch {et~al.}(2006)Frisch, Sampoorna, \&
  Nagendra}]{frisch_stochastic_2006}
Frisch, H., Sampoorna, M., \& Nagendra, K.~N. 2006, Astronomy and Astrophysics,
  453, 1095

\bibitem[{Graham {et~al.}(2009)Graham, Danilovic, \&
  Sch{\"u}ssler}]{pietarila_graham_turbulent_2009}
Graham, J.~P., Danilovic, S., \& Sch{\"u}ssler, M. 2009, The Astrophysical
  Journal, 693, 1728

\bibitem[{Knobloch \& Rosner(1981)}]{knobloch_spectrum_1981}
Knobloch, E. \& Rosner, R. 1981, The Astrophysical Journal, 247, 300

\bibitem[{Kosugi {et~al.}(2007)Kosugi, Matsuzaki, Sakao, Shimizu, Sone,
  Tachikawa, Hashimoto, Minesugi, Ohnishi, Yamada, Tsuneta, Hara, Ichimoto,
  Suematsu, Shimojo, Watanabe, Shimada, Davis, Hill, Owens, Title, Culhane,
  Harra, Doschek, \& Golub}]{kosugi_hinode_2007}
Kosugi, T., Matsuzaki, K., Sakao, T., {et~al.} 2007, Solar Physics, 243, 3

\bibitem[{Lites(2002)}]{lites_characterization_2002}
Lites, B.~W. 2002, The Astrophysical Journal, 573, 431

\bibitem[{{L{\'o}pez Ariste}(2002)}]{lopez_ariste_asymmetry_2002}
{L{\'o}pez Ariste}, A. 2002, The Astrophysical Journal, 564, 379

\bibitem[{{L{\'o}pez Ariste} {et~al.}(2007){L{\'o}pez Ariste}, Malherbe, {Manso
  Sainz}, {Asensio Ramos}, {Ram{\'i}rez V{\'e}lez}, \& {Mart{\'i}nez
  Gonz{\'a}lez}}]{lopez_ariste_turbulent_2007}
{L{\'o}pez Ariste}, A., Malherbe, J.~M., {Manso Sainz}, R., {et~al.} 2007, in
  SF2A-2007: Proceedings of the Annual meeting of the French Society of
  Astronomy and Astrophysics, ed. J.~Bouvier, A.~Chalabaev, \& C.~Charbonnel,
  592

\bibitem[{{Mart{\'i}nez Gonz{\'a}lez} {et~al.}(2008{\natexlab{a}}){Mart{\'i}nez
  Gonz{\'a}lez}, {Asensio Ramos}, Carroll, Kopf, {Ram{\'i}rez V{\'e}lez}, \&
  Semel}]{martinez_gonzalez_pca_2008}
{Mart{\'i}nez Gonz{\'a}lez}, M.~J., {Asensio Ramos}, A., Carroll, T.~A.,
  {et~al.} 2008{\natexlab{a}}, Astronomy and Astrophysics, 486, 637

\bibitem[{{Mart{\'i}nez Gonz{\'a}lez} {et~al.}(2008{\natexlab{b}}){Mart{\'i}nez
  Gonz{\'a}lez}, {Asensio Ramos}, {L{\'o}pez Ariste}, \& {Manso
  Sainz}}]{martinez_gonzalez_near-ir_2008}
{Mart{\'i}nez Gonz{\'a}lez}, M.~J., {Asensio Ramos}, A., {L{\'o}pez Ariste},
  A., \& {Manso Sainz}, R. 2008{\natexlab{b}}, Astronomy and Astrophysics, 479,
  229

\bibitem[{{Mart{\'i}nez Gonz{\'a}lez} {et~al.}(2010){Mart{\'i}nez
  Gonz{\'a}lez}, {Manso Sainz}, {Asensio Ramos}, {L{\'o}pez Ariste}, \&
  Bianda}]{martinez_gonzalez_statistical_2010}
{Mart{\'i}nez Gonz{\'a}lez}, M.~J., {Manso Sainz}, R., {Asensio Ramos}, A.,
  {L{\'o}pez Ariste}, A., \& Bianda, M. 2010, The Astrophysical Journal, 711,
  L57

\bibitem[{{Mart{\'i}nez Pillet} {et~al.}(1997){Mart{\'i}nez Pillet}, Lites, \&
  Skumanich}]{martinez_pillet_active_1997}
{Mart{\'i}nez Pillet}, V., Lites, B.~W., \& Skumanich, A. 1997, The
  Astrophysical Journal, 474, 810

\bibitem[{Mihalas(1978)}]{mihalas_stellar_1978}
Mihalas, D. 1978, Stellar atmospheres /2nd edition/, 2nd edn. (San Francisco:
  W.H. Freeman and Co.)

\bibitem[{Nakagawa \& Priest(1973)}]{nakagawa_energy_1973}
Nakagawa, Y. \& Priest, E.~R. 1973, The Astrophysical Journal, 179, 949

\bibitem[{Nordlund {et~al.}(1992)Nordlund, Brandenburg, Jennings, Rieutord,
  Ruokolainen, Stein, \& Tuominen}]{nordlund_dynamo_1992}
Nordlund, A., Brandenburg, A., Jennings, R.~L., {et~al.} 1992, The
  Astrophysical Journal, 392, 647

\bibitem[{Parker(1982)}]{parker_dynamics_1982}
Parker, E.~N. 1982, The Astrophysical Journal, 256, 292

\bibitem[{Petrovay(2001)}]{petrovay_turbulence_2001}
Petrovay, K. 2001, Space Science Reviews, 95, 9

\bibitem[{{Sainz Dalda} {et~al.}(2012){Sainz Dalda}, Mart{\'i}nez-Sykora,
  {Bellot Rubio}, \& Title}]{SainzDalda12}
{Sainz Dalda}, A., Mart{\'i}nez-Sykora, J., {Bellot Rubio}, L., \& Title, A.
  2012, The Astrophysical Journal, in Press

\bibitem[{Sampoorna {et~al.}(2008)Sampoorna, Nagendra, Frisch, \&
  Stenflo}]{sampoorna_zeeman_2008}
Sampoorna, M., Nagendra, K.~N., Frisch, H., \& Stenflo, J.~O. 2008, Astronomy
  and Astrophysics, 485, 275

\bibitem[{{S{\'a}nchez Almeida}(2007)}]{sanchez_almeida_simple_2007}
{S{\'a}nchez Almeida}, J. 2007, The Astrophysical Journal, 657, 1150

\bibitem[{{S{\'a}nchez Almeida} \&
  Lites(1992)}]{sanchez_almeida_observation_1992}
{S{\'a}nchez Almeida}, J. \& Lites, B.~W. 1992, The Astrophysical Journal, 398,
  359

\bibitem[{{Sanchez Almeida} {et~al.}(1996){Sanchez Almeida}, {Ruiz Cobo}, \&
  del {Toro Iniesta}}]{sanchez_almeida_heights_1996}
{Sanchez Almeida}, J., {Ruiz Cobo}, B., \& del {Toro Iniesta}, J.~C. 1996,
  Astronomy and Astrophysics, 314, 295

\bibitem[{Solanki \& Pahlke(1988)}]{solanki_can_1988}
Solanki, S.~K. \& Pahlke, K.~D. 1988, Astronomy and Astrophysics, 201, 143

\bibitem[{Solanki \& Stenflo(1984)}]{solanki_properties_1984}
Solanki, S.~K. \& Stenflo, J.~O. 1984, Astronomy and Astrophysics, 140, 185

\bibitem[{Stenflo(1973)}]{stenflo_magnetic-field_1973}
Stenflo, J.~O. 1973, Solar Physics, 32, 41

\bibitem[{Stenflo(2010)}]{stenflo_distribution_2010}
Stenflo, J.~O. 2010, Astronomy and Astrophysics, 517, 37

\bibitem[{{Trujillo Bueno} {et~al.}(2004){Trujillo Bueno}, Shchukina, \&
  {Asensio Ramos}}]{trujillo_bueno_substantial_2004}
{Trujillo Bueno}, J., Shchukina, N., \& {Asensio Ramos}, A. 2004, Nature, 430,
  326

\bibitem[{Tsuneta {et~al.}(2008)Tsuneta, Ichimoto, Katsukawa, Nagata, Otsubo,
  Shimizu, Suematsu, Nakagiri, Noguchi, Tarbell, Title, Shine, Rosenberg,
  Hoffmann, Jurcevich, Kushner, Levay, Lites, Elmore, Matsushita, Kawaguchi,
  Saito, Mikami, Hill, \& Owens}]{tsuneta_solar_2008}
Tsuneta, S., Ichimoto, K., Katsukawa, Y., {et~al.} 2008, Solar Physics, 249,
  167

\bibitem[{Viticchi{\'e} {et~al.}(2010)Viticchi{\'e}, {S{\'a}nchez Almeida},
  {Del Moro}, \& Berrilli}]{viticchie_interpretation_2010}
Viticchi{\'e}, B., {S{\'a}nchez Almeida}, J., {Del Moro}, D., \& Berrilli, F.
  2010, Interpretation of {HINODE} {SOT/SP} asymmetric Stokes profiles observed
  in quiet Sun network and internetwork,
  {http://adsabs.harvard.edu/abs/2010arXiv1009.6065V}

\end{thebibliography}

\end{document}